\documentclass[12pt]{article}
\usepackage{epsf}

\newcommand{\beq}{\begin{equation}}
\newcommand{\eeq}{\end{equation}}
\newcommand{\beqa}{\begin{eqnarray}}
\newcommand{\eeqa}{\end{eqnarray}}

\def\barD{\overline D{}^0}

\def\barH{\overline H{}^0}

\def\DDbar{D^0-\overline D{}^0}

\def\am{a_m}

\def\be{\begin{equation}}
\def\ee{\end{equation}}
\def\bea{\begin{eqnarray}}
\def\eea{\end{eqnarray}}

\def\lsim{\mathrel{\lower4pt\hbox{$\sim$}}\hskip-12pt\raise1.6pt\hbox{$<$}\;
}

\def\xba{\bar}

\begin{document}

\begin{titlepage}

\begin{flushright}
WSU-HEP-0102\\
AMES-HET 02-06\\
SLAC-PUB-9292\\
{\tt hep-ph/0207165}\\[0.2cm]
July 2002\\
\end{flushright}

\vspace{1.5cm}
\begin{center}
\Large\bf\boldmath
Lifetime differences in heavy mesons
with time independent measurements
\unboldmath
\end{center}

\vspace{1.0cm}
\begin{center}
David Atwood$^1$\\[0.1cm]
{\sl  Department of Physics and Astronomy, Iowa State University \\
Ames, IA 50011} \\[0.5cm]
Alexey A. Petrov$^2$\\[0.1cm]
{\sl Department of Physics and Astronomy, Wayne State University\\
Detroit, MI 48201}
\end{center}

\footnotetext[1]{email: atwood@iastate.edu}
\footnotetext[2]{email: apetrov@physics.wayne.edu}

\vspace{0.8cm}
\begin{abstract}
\vspace{0.2cm}
\noindent Heavy meson pairs produced in the decays of heavy quarkonium
resonances at $e^+e^-$ machines (beauty and tau-charm factories) have the
useful property that the two mesons are in the CP-correlated states.
By tagging one of the mesons as a CP eigenstate, a lifetime difference
of heavy neutral meson mass eigenstates 
$\Delta\Gamma$ may be determined by measuring the leptonic
branching ratio of the other meson. We discuss the use of this and related methods 
both in the case where time dependent mixing is small and when it is
significant. We consider the impact of possible CP-violating effects
and present the complete results for CP-entangled decay rates with CP-violation
taken into account.
\end{abstract}

\end{titlepage}

\section{Introduction}

One of the most important motivations for studies
of heavy meson mixing is the possibility of observing a
signal from new physics which can be separated from the one 
generated by the Standard Model (SM) interactions. 
If $H^0$ is a neutral heavy meson, $\Delta Q=2$ transitions, 
occurring only at one loop in the Standard Model, as well as possible 
new interactions, generate contributions to the effective
operators that change $H^0$ state into $\barH$ state
leading to the mass eigenstates

\begin{equation} \label{definition1}
| H_{^1_2} \rangle =
p | H^0 \rangle \pm q | \bar H^0 \rangle,
\end{equation}

\noindent
where the complex parameters $p$ and $q$ are 
obtained from diagonalizing the $H^0-\barH$ mass matrix~\cite{Donoghue:1992dd}. 
It is conventional to parameterize the mass and width splittings between
these eigenstates by

\begin{eqnarray} \label{definition}
x \equiv \frac{m_2-m_1}{\Gamma}, ~~
y \equiv \frac{\Gamma_2 - \Gamma_1}{2 \Gamma},
\end{eqnarray}

\noindent
where $m_{1,2}$ and $\Gamma_{1,2}$ are the masses and widths of
$H_{1,2}$  and the mean width and mass are 
$\Gamma=(\Gamma_1+\Gamma_2)/2$ and $m=(m_1+m_2)/2$. 
Since $y$ is constructed from the decays of $H$ into physical states, 
it should be dominated by the Standard Model contributions, unless 
new physics significantly modifies $\Delta Q=1$ interactions.

Our goal in this paper is to discuss time independent methods to determine
$y$ for the various oscillating hadron species (i.e. $H=D^0$, $B^0$ and
$B_s$) at electron-positron colliders where an $H\barH$ pair can be
produced in a correlated initial state. In particular we will consider
methods which are time independent and therefore may be applied in
experiments where the time history is difficult to obtain.
Results which are linear, rather then quadratic, in $y$ are of particular 
interest because $y$ is generically small for $B_d^0$ and $D^0$ mesons. 

%
%
%
%
%
%

Indeed, in the case of $D^0$, such studies may be carried out at tau-charm
factories running on the $\psi(3370)$ resonance. In this case $x$ is small
and it could well be that $y\gg x$. It has been argued that the Standard
Model $y>x$ \cite{Falk:2001hx} with $y\sim O(1\%)$. Thus, lifetime
differences may be the dominant form of mixing in $D^0$ and their study is
perhaps within the reach of proposed experiments.

In the case of $B_d^0$, $x=0.730 \pm 0.029$~\cite{pdb2000}
and $y$ is small, of the order of $0.3\%$ \cite{Dighe:2001gc}. 
It is possible that at future B-factories there would be
enough statistics to measure $y$. In addition, if $e^+e^-$ machines are
run at the $\Upsilon(5s)$ resonance, these methods could be used to
investigate $y$ in the $B_s$. In the $B_s$ system, it has been estimated
that $y$ may be particularly large (5-15\%)~\cite{bs_est} and indeed
similar methods have been previously discussed in~\cite{Atwood:2001js}.

In the cases of $B_s$ and $D^0$, it is thought that CP violation in the
mixing is small while in $B^0$, mixing is known to produce large CP
violating effects~\cite{babar,belle}. In our methods, CP violation in
mixing will impact the signal, generally reducing it,  
so we will derive the formalism in a framework which includes CP violation.

The paper will proceed as follows.  In Section 2 we will discuss the
formalism which applies in the case where CP violation is negligible. In
Section~3 we will discuss the application to $D^0{\barD}$ production at
a charm factory assuming there is no CP violation in $D^0$ oscillation or
decay.  In Section~4 we will generalize the formalism to the case where CP
violation is present and consider the application to $B$ and $D$ mesons.  
In Section~5 we will give our conclusions.
In the appendix we give the time integrated correlated decay rates for
$H^0H^0$ decaying to various combinations of final states where indirect
CP violation is present.

\section{Formalism if CP is Conserved}

The essential point of our method is best illustrated
in cases where $CP$ violation may be neglected. 
In such cases, $p=q$, so mass eigenstates also
become eigenstates of $CP$, which we denote: 
\begin{eqnarray} \label{defCP}
| H_{\pm} \rangle =
\frac{1}{\sqrt{2}} \left[
| H^0 \rangle \pm | \bar H^0 \rangle \right].
\end{eqnarray}
The crucial point here is that these CP eigenstates
$| H_{\pm} \rangle$ do not evolve with time.
We can take advantage of this fact at 
threshold $e^+e^-$ machines, such as BaBar or CLEO-c. 
$H^0\xba H^0$ production at these machines is through 
resonances leading to $HH$ pairs being in a quantum mechanically 
coherent state. Thus, if the production resonance has angular momentum
$L$, the quantum mechanical state at the time of $H\barH$ production is
\begin{eqnarray}
\Psi_L\equiv
|H^0 \barH \rangle_L = \frac{1}{\sqrt{2}}
\left \{
| H^0 (k_1)\barH (k_2) \rangle +
(-1)^L | H^0 (k_2)\barH (k_1) \rangle
\right \}.
\label{entangle1}
\end{eqnarray}
where $k_1$ and $k_2$ are the momenta of the mesons. 
Rewriting this in terms of the CP basis we arrive at
\begin{eqnarray} 
\Psi_{2n} &=& \frac{1}{\sqrt{2}}
\left \{
| H_+ (k_1) H_+ (k_2) \rangle 
-
| H_- (k_1) H_- (k_2) \rangle 
\right \}
\nonumber\\
\Psi_{2n+1} &=& \frac{1}{\sqrt{2}}
\left \{
| H_- (k_1) H_+ (k_2) \rangle 
+
| H_+ (k_1) H_- (k_2) \rangle 
\right \}
\label{entangle2}
\end{eqnarray}
Thus in the $L=$ odd case, which would apply to the experimentally
important $\psi(3770)$ and $\Upsilon(4S)$ resonances, the CP eigenstates
of the $H$ mesons are anti-correlated while if $L=$ even the eigenstates
are correlated\footnote{While $L=$ even resonances are not directly produced in 
$e^+e^-$ collisions, quantum-mechanically {\it symmetric} states 
can be produced 
in the decays, such as $\psi(4140) \to D {\overline D} \gamma$. 
In the following, 
$L=$ even case can also refer to this situation.}. In either case the
correlation 
between the eigenstates is independent of when they decay. In this way, 
if meson $H(k_1)$ decays to the final state which is also a CP eigenstate, 
then the CP eigenvalue of the meson $H(k_2)$ can be determined. 
 
Using this eigenstate correlation as a tool to investigate CP violation
has been suggested by~\cite{Falk:2000ga}\footnote{For other measurements that 
involve CP correlations to study CP violation in $D$-mesons see~\cite{Azimov}.}. 
In this paper we will take advantage of such correlations for the experimental 
investigation of lifetime differences. The idea is fairly straightforward: we look 
at decays of the form $\psi_L\to (H\to S_\sigma)(H\to Xl\nu)$ where $S_\sigma$ is a CP
eigenstate of eigenvalue $\sigma=\pm 1$ and $\psi_L$ generically means any
resonance of angular momentum $L$ that decays to $H\barH$ (see also a footnote
after the Eqn.~(\ref{entangle2}).  

In this scenario, the CP quantum numbers of the $H(k_2)$ is thus determined.
The semi-leptonic {\it width} of this meson should be independent of the CP 
quantum number since it is flavor specific.
It follows that the semi-leptonic {\it branching ratio} of
$H(k_2)$ will be inversely proportional to the total width of that meson.
Since we know whether $H(k_2)$ is a $H_+$ or and $H_-$ from the decay of
$H(k_1)$, we can easily determine $y$ in terms of the semileptonic
branching ratios of $H_\pm$. 

This can be expressed simply by introducing the ratio
\beq \label{DefCor}
R^L_\sigma=
\frac{\Gamma[\psi_L \to (H \to S_\sigma)(H \to X l^\pm \nu )]}{
\Gamma[\psi_L \to (H \to S_\sigma)(H \to X)]~Br(H^0 \to X l \nu)},
\eeq
where $X$ in $H \to X$ stands for an inclusive set of all
final states. A deviation from $R^L_\sigma=1$ implies a 
lifetime difference. From this experimentally obtained quantity, 
we extract $y$ by
\begin{eqnarray}
R^L_\sigma=
{1\over 1 + (-1)^L\sigma y},
~~~~~~
y=\sigma (-1)^L \frac{R^L_\sigma-1}{R^L_\sigma}
\label{DefCor2}
\end{eqnarray}

\section{Charmed Mesons if CP is Conserved}

Let us consider now the production of $D^0\barD$ mesons at an electron
positron collider. In the Standard Model, CP violation is expected to be
small in $D^0$ hence the above formalism should apply directly to this
case.  For instance, the new tau-charm factory, under construction at
CESR, will allow for simple and effective CP tagging in the case of
$\psi(3770)\to D{\overline D}$ because there are numerous candidates for
$S_\sigma$ in $D^0$ decay which have branching ratios in the few percent
range, for instance $K_S\pi^0$ (1.05\%);  $K_S\omega$ (1.05\%);  
$K_S\eta^\prime$ (0.85\%);  $\pi^+\pi^-$ (0.15\%). In additon, the modes
$K^{*0}\pi^0$ and $K^{*0}\rho^0$ may be used provided the $K^{*0}$ itself
decays to a CP eigenstate, $K_{S,L}\pi^0$ and one can separate the main
amplitude from cross channel processes. Similar comments apply to analogous
states containing higher neutral kaonic resonances. 

%
%

These modes are thus candidates for $S_\sigma$ in Eqn.~(\ref{DefCor}). 
If we write Eqn.~(\ref{DefCor})  in terms of the semi-leptonic
branching ratio of $D_\pm$, ${\cal B}_\pm^\ell$, then equation
Eqn.~(\ref{DefCor2})  becomes:

\begin{eqnarray}
\left({{\cal B}^\ell_+(D)\over {\cal
B}^\ell_-(D)}
-{{\cal B}^\ell_-(D)\over {\cal B}^\ell_+(D)}\right) = 4y.
\label{main_eqn}
\end{eqnarray} 

\noindent
In either case the statistical uncertainty in $y$ is given by 
\begin{eqnarray}
\Delta y = \left (2 N_0 {\cal A}^\ell {\cal B}^\ell 
{\cal A}^\sigma {\cal B}^\sigma \right)^{-{1\over 2}}  
\label{delta_y_1}
\end{eqnarray}
where $N_0$ is the initial number of $\psi$'s, ${\cal A}^\sigma$
and ${\cal A}^\ell$ are the acceptances for the CP eigenstate modes and
the semileptonic modes respectively while ${\cal B}^\sigma$ and ${\cal
B}^\ell$ are the branching ratios for those modes.  In general, of course,
we can combine the statistics for a number of modes so, as an example, if
we assume that ${\cal B}^\ell=12\%$, ${\cal B}^\sigma=2\%$, with ${\cal
A}^\ell{\cal A}^\sigma=0.1$ then $N_0=10^8$ gives $\Delta
y=0.5\%$.

At present, the information about the $\DDbar$ mixing parameters $x$ and
$y$ comes from the time-dependent analyses that can roughly be divided
into two categories. First, more traditional analyses study time
dependence of $D \to f$ decays, where $f$ is the final state that can be
used to tag the flavor of the decayed meson. The most popular is the
non-leptonic doubly Cabibbo suppressed (DCS) decay $D \to K^+ \pi^-$.  
Time-dependent studies allow one to separate the doubly Cabibbo suppressed
decay from the mixing contribution,

\begin{eqnarray}\label{Kpi} 
\Gamma[D^0(t) \to K^+ \pi^-]
=e^{-\Gamma t}|A_{K^-\pi^+}|^2 \qquad\qquad\qquad\qquad\qquad
\qquad
\nonumber \\
\times \left[
R+\sqrt{R}R_m(y'\cos\phi-x'\sin\phi)\Gamma t
+\frac{R_m^2}{4}(y^2+x^2)(\Gamma t)^2
\right],
\end{eqnarray}

\noindent where $R$ is the ratio of Cabibbo favored (CF) and doubly
Cabibbo suppressed decay rates and $R_m=|p/q|$ while $\phi=arg(p/q)$.
Since $x$ and $y$ are small, the best constraint comes from the linear
terms in $t$ that are also linear in $x$ and $y$.  Using this method,
direct extraction of $x$ and $y$ is not possible from Eq.~(\ref{Kpi}) due
to unknown relative strong phase $\delta$ of DCS and CF amplitudes (see
\cite{Falk:1999ts} for extensive discussion), as
$x'=x\cos\delta+y\sin\delta$, $y'=y\cos\delta-x\sin\delta$. This phase,
however, can be measured independently~\cite{GGR}. The corresponding
formula can also be written for $\barD$ decay with $x' \to -x'$ and $R_m
\to R_m^{-1}$ \cite{Bergmann:2000id}.

Another
method to measure $D^0$ mixing is to  
compare the lifetimes extracted from the analysis
of $D$ decays into the CP-even and CP-odd final states. This study is also
sensitive to a linear function of $y$, via
\beq
\frac{\tau(D \to K^-\pi^+)}{\tau(D \to K^+K^-)}-1=
y \cos \phi + x \sin \phi \left[\frac{1-R_m^2}{2}\right].
\eeq
Time-integrated studies of the semileptonic transitions are sensitive 
to the quadratic form $x^2+y^2$ and at the moment are not competitive with the
analyses discussed above.

\section{Formalism if CP is Violated in $H^0\barH$ Oscillations}

In the case where CP violation is present in the $H^0\barH$ mixing, it is
necessary to consider general time dependent entangled states of the
$H^0\barH$ pair.  
Following the notation of \cite{Donoghue:1992dd}, we will denote the
wave function $|H(t)\rangle$ at a given moment in time, $t$, by a two
element vector:
\begin{eqnarray}
|H(t)\rangle = a |H^0\rangle+\overline a |\barH\rangle\equiv 
\left (
\begin{array}{c}
a\\
\overline a\\
\end{array}
\right)
\end{eqnarray}

\noindent
CPT conservation forces the general mass matrix in the following form

\begin{eqnarray}
{\cal M}\equiv \hat M+i\hat \Gamma/2 = 
\left (
\begin{array}{cc}
A &p^2\\
q^2&A
\end{array}
\right)
\end{eqnarray}

\noindent 
where $\hat M$ and $\hat \Gamma$ are Hermitian while $A$, $p$ and $q$ are
in general complex numbers. The effects of CP violation in the
system are usually parameterized in terms of the ratio:

\begin{eqnarray} \label{pq}
{p\over q}\equiv {1+\epsilon\over 1-\epsilon}
\equiv 
R_m e^{i\phi}
\end{eqnarray}

\noindent
In the limit of CP conservation in mixing matrix, $R_m=1$.
Even if CP is violated, in the case of heavy neutral mesons, it is
expected that
$R_m\approx 1$. 
The phase $\phi$, of course,
depends on the convention one uses for weak phases that can be traded off
against the weak phase in the decay in the usual way. In our discussion it
will be useful to assume that we are using a convention where we 
have absorbed any weak phase from the decay into the the mixing.

%
%
%
%

For an isolated $H$ meson, the wave function at time $t$ is related to the
the wave function at $t=0$ by:
\begin{eqnarray}
|H(t)\rangle=U_t|H(0)\rangle,
\end{eqnarray}
where the time evolution operator $U_t$ satisfies the equation
\begin{eqnarray}
i{d U_t\over dt}={\cal M} U_t,~~~~U_0=1.
\end{eqnarray}
This Schr\"odinger-like equation can be solved to yield the 
familiar result
\begin{eqnarray}
U_t=
\left (
\begin{array}{cc}
g_+(t)  & (p/q)~g_-(t)\\
(q/p)~g_-(t) & g_+(t)
\end{array}
\right ).
\end{eqnarray}
Here, the time dependence of $D^0$ and $\barD$ is driven by 
\begin{eqnarray}
g_+(t)&=&\left(
\cosh y\tau/2\cos x\tau/2
-i\sinh y\tau/2\sin x\tau/2
\right) e^{-\mu\tau/2} 
\nonumber\\
g_-(t)&=&\left(
-\sinh y\tau/2\cos x\tau/2
+i\cosh y\tau/2\sin x\tau/2
\right) e^{-\mu\tau/2} 
\end{eqnarray}
with $\tau=\Gamma t$ and $\mu=1+2i m/\Gamma$.

Let us now consider the time integrated decay rate for a single $H$ to a
final state $f$.  If $a$ and $\overline a$ are the amplitudes for $H^0$
and $\barH$ to decay to $f$ respectively and $|\psi_0 \rangle$ is the
initial wave function for the meson, then the time integrated decay rate
is
\begin{eqnarray}
2\Gamma_f(\rho_0)
=
(Q+P)  tr[\rho_f\rho_0]
+(Q-P) tr[v^\dag \rho_f\rho_0]
-2 Re \left [ (yQ-ixP) tr[\rho_f v\rho_0]    \right ],
\end{eqnarray}
where 
\begin{eqnarray} \label{rhos}
\rho_0=|\psi_0\rangle\langle\psi_0|
~~~~
\rho_f=\left (
\begin{array}{cc}
|a|^2&a^*\overline a\\
{\overline a}^* a&|{\overline a}|
\end{array}
\right) 
~~~~
v=\left (
\begin{array}{cc}
0 & p/q\\
q/p & 0
\end{array}
\right)
\end{eqnarray}
and
\begin{eqnarray}
P=1/(1+x^2) ~~~~ Q=1/(1-y^2).
\end{eqnarray}
Of particular interest is the case case where $f$ is a CP eigenstate with
CP=$\sigma=\pm 1$. If we assume $R_m=1$, as would be the case for
$B_d\to \psi K_s$ in the Standard Model, then
\begin{eqnarray}
{1\over 2}\left ({\Gamma_f(H^0)+\Gamma_f(\barH)}\right )
&=&
{1-\sigma y\cos\phi\over 1-y^2}\Gamma_0
\label{bsum}\\
{1\over 2}\left ({\Gamma_f(H^0)-\Gamma_f(\barH)}\right )
&=&
\sigma{x\sin\phi\over 1+x^2}\Gamma_0,
\label{bdiff}
\end{eqnarray}
where $\Gamma_0$ is the decay rate for $H$ to $f$. 

This can be easily generalized to the case of the entangled initial state 
which presents itself in the creation of $H^0\barH$ pairs from a $\Psi_L$ 
resonance.  As was shown in Eqns.~(\ref{entangle1},\ref{entangle2}), the 
states of interest
can be decomposed into the coherent sum of products of flavor (or CP)
eigenstates.  Using Eqn.~(\ref{entangle1}) we can write the time
integrated correlated decay rate for $\Psi_L\to (H\to f_1)(H\to f_2)$ is:
\begin{eqnarray}
\Gamma^{f_1f_2}(\Psi_L)
=
\Gamma_{f_1}(H^0)\Gamma_{f_2}(\barH)
+\Gamma_{f_1}(\barH)\Gamma_{f_2}(H^0)
\nonumber\\
+(-1)^L\left [
\Gamma_{f_1}(\rho_{+-})\Gamma_{f_2}(\rho_{-+})+
\Gamma_{f_1}(\rho_{-+})\Gamma_{f_2}(\rho_{+-})
\right]
\label{master_eqn}
\end{eqnarray}
where $\rho_{ik}$ are the matrices:
\begin{eqnarray}
&&\rho_{++}=\left ( \begin{array}{cc}1&0\\ 0&0\end{array} \right )
~~~
\rho_{--}=\left ( \begin{array}{cc}0&0\\ 0&1\end{array} \right )
\nonumber \\
&&\rho_{+-}=\left ( \begin{array}{cc}0&1\\ 0&0\end{array} \right )
~~~
\rho_{-+}=\left ( \begin{array}{cc}0&0\\ 1&0\end{array} \right )
\label{rho_def}
\end{eqnarray}
Let us return to the calculation of $y$ through the
determination of $R^L_\sigma$ defined in 
Eqn.~(\ref{DefCor}). In the case $L=1$ the numerator is
$\Gamma^{S_\sigma X_l}(\Psi_1)$ and so
Eqn.~(\ref{master_eqn}) implies:
\begin{eqnarray}
\Gamma^{S_\sigma X_l}(\Psi_L)
\propto
{
2+x^2(R_m^{\pm 2}+1)+y^2(R_m^{\pm 2}-1)
\over
2(1+x^2)(1-y^2)
}
\label{gam_l_sig}
\end{eqnarray}
where we assume there is no further CP violation in the
decay amplitude, and the $\pm$ signs are for the positively 
and negatively charged leptons respectively.

The denominator of Eqn.~(\ref{DefCor}) is given by Eqn.~(\ref{master_eqn})
where $f_1$ is $S_\sigma$ and all possible values of 
$f_2$ are summed over. In this case the $L$ dependent terms
vanish and the rest simplifies to:
\begin{eqnarray}
\Gamma^{f_1X}(\Psi_L)
\equiv \sum_{f_2}
\Gamma^{f_1f_2}(\Psi_L)
=\Gamma_{f_1}({\bf 1})/2,
\end{eqnarray}
where ${\bf 1}$ is the identity matrix. 
This is proportional to the
average between the time integrated decay rate of $H^0$ and 
$\barH$ to the final state $f_1$. It is easiest to see that in the 
CP-eigenstate basis spanned by the states of Eq.~(\ref{entangle2}).
Indeed, in the particular example of $L=$ odd the time-integrated decay 
rate is
\begin{eqnarray} \label{master_eqn2}
\Gamma^{f_1X}(\Psi_1)
= \int d t_1 d t_2 ~\frac{1}{2} ~\biggl[
~Tr~\left[~U_{t2}^\dagger ~\rho_X ~U_{t2} ~\rho_{++} \right]~
Tr~\left[~U_{t1}^\dagger ~\rho_{f1} ~U_{t1} ~\rho_{--} \right]&&
\nonumber \\
+~
Tr~\left[~U_{t2}^\dagger ~\rho_X ~U_{t2} ~\rho_{--} \right]~
Tr~\left[~U_{t1}^\dagger ~\rho_{f1} ~U_{t1} ~\rho_{++} \right]&&
\nonumber \\
-~
Tr~\left[~U_{t2}^\dagger ~\rho_X ~U_{t2} ~\rho_{+-} \right]~
Tr~\left[~U_{t1}^\dagger ~\rho_{f1} ~U_{t1} ~\rho_{-+} \right]&&
\\
-~
Tr~\left[~U_{t2}^\dagger ~\rho_X ~U_{t2} ~\rho_{-+} \right]
Tr~\left[~U_{t1}^\dagger ~\rho_{f1} ~U_{t1} ~\rho_{+-} \right]&&
\biggr],
\nonumber
\end{eqnarray}
where $\rho_X$ and $\rho_{f1}$ are the matrices for $H\to X$ and $H\to f_1$
decay amplitudes respectively (see Eqn~(\ref{rhos})).
It is easy to see that in the {\it mass eigenstate} basis
$U_{t2}^\dagger ~\rho_X ~U_{t2}|_{mass} =
diag(\Gamma_1 e^{-\Gamma_1 t_2}, \Gamma_2 e^{-\Gamma_2 t_2})$.
In principle, this needs to be translated back to the CP eigenstate basis.
However, integration with respect to $t_2$ yields a unit matrix, which is 
invariant under the change of basis.
This simplifies the Eq.~(\ref{master_eqn2}) considerably, which, after
taking the corresponding traces transforms into

\begin{eqnarray} 
\Gamma^{f_1X}(\Psi_1)
= \frac{1}{2} \int d t ~
Tr~\left[~\rho_{f1} ~U_{t} ~U_{t}^\dagger \right]
\biggr],
\nonumber
\end{eqnarray}

\noindent
which, in the limit $R_m=1$, becomes for the semileptonic final state 
(a complete expression is available in the Appendix)

\begin{eqnarray}
\Gamma^{X X_l}(\Psi_L)
\propto
{
1+(-1)^L\cos\phi
\over
(1-y^2).
}
\label{gam_X_sig}
\end{eqnarray}
In the limit $R_m=1$ the ratio of Eqs.~(\ref{gam_l_sig}) and 
(\ref{gam_X_sig}) becomes:
\begin{eqnarray}
R_\sigma^L={1\over 1+(-1)^L\sigma y \cos\phi}
\end{eqnarray}
In which case the generalization of Eqn.~(\ref{DefCor2}) is 
\begin{eqnarray}
y\cos\phi=
(-1)^L {\sigma}
{R^L_\sigma-1\over R^L_\sigma}
\label{y-cos-phi}.
\end{eqnarray}
So we can regard the measurement of $R^L_\sigma$ as leading to a
determination of $y\cos\phi$. A similar result holds for the non-leptonic
final state (such as $D \to K\pi$, with corrections proportional to $R$).
In the case where $R_m\neq 1$ the corresponding expression depends on $x$
as well as $y$. For instance, for $L=1$
\begin{eqnarray}
R_\sigma^L={PQ(1+R_mx^2\cosh\am+R_my^2\sinh\am)\over
(Q\cosh^2\am-P\sinh^2\am-xP\sinh\am\sin\phi-yQ\cosh\am\cos\phi)}
\end{eqnarray}


\noindent
where $\am=\log(R_m)=\log\sqrt{1+A_m}\approx A_m/2$  
\cite{Bergmann:2000id}.
Expanding this to first order in $\am$ we obtain:

\begin{eqnarray}
R_\sigma^L={1\over 1-\sigma y \cos\phi}
+
{(x^2+y^2)(1-y\cos\phi) +x(1-y^2)\sin\phi
\over
(1-y\cos\phi)^2(1+x^2)
}\am
+O(\am^2)
\end{eqnarray}

\noindent
Thus, if we define $\hat y$ by

\begin{eqnarray}
\hat y \cos\phi =\sigma{R^L_\sigma - 1 \over R^L_\sigma}
\end{eqnarray}

\noindent
If we expand $\hat y$ to first order in $\am$ we obtain:

\begin{eqnarray}
\hat y
=
y 
-\am
\left [
{
(x^2+y^2)(1-y\cos\phi)+(1-y^2)x\sin\phi)
\over
(1+x^2)\cos\phi
}
\right ]
\end{eqnarray}

\noindent
Clearly then, Eqn.~(\ref{y-cos-phi})
gives $y$ only if $\am$ is known to be small.
The actual value of $\am$ can be experimentally obtained from 
the semileptonic decay asymmetry~\cite{pdb2000}.

In our discussion we will now assume that $R_m\approx 1$ 
and so the ratio $R^L_\sigma$ gives
us $y\cos\phi$ through Eqn.~(\ref{y-cos-phi}). The error in determining
$y$ is thus given by the generalization of Eqn.~(\ref{delta_y_1})

\begin{eqnarray}
\Delta y \cos\phi= \left (2 N_0 {\cal A}^\ell {\cal B}^\ell 
{\cal A}^\sigma {\cal B}^\sigma \right)^{-{1\over 2}}  
\label{delta_y_2}
\end{eqnarray}

\noindent
In the case of $D^0$, the systematics for $\Delta y\cos\phi$ is the same
as the systematics for $\Delta y$ in the CP conserving case discussed
above.

In the case of $B^0$, $\phi$ which is equal to $2\beta$ in the Standard
Model has been measured at the BaBar and BELLE
experiments~\cite{babar,belle} The average of these two results is
currently $\sin 2\beta=0.78\pm 0.08$ thus $\cos 2\beta \approx 0.6$.  
Thus, if we take $N_0=10^8$ and use only $\psi K_S$ decay mode with
$\psi\to l^+l^-$ and assume that ${\cal A}_\sigma{\cal A}_l\approx 1/4$
then $\Delta y\cos 2\beta = 0.06$ corresponding to $\Delta y = 0.1$.
Clearly bringing in additional $S_\sigma$ modes will improve determination 
of $\Delta y$.
We can also improve the statistics by using flavor specific decays of the
$B^0$ other than pure leptonic decays. The BaBaR and Belle experiments
have made considerable progress in their ability to accomplish this and
obtain an effective value of ${\cal A}_lB_l\approx 0.7$. Using this
result, the above gives $\Delta y \approx 0.026$.

To produce correlated $B_s$ pairs one needs to run an electron-positron
machine at the $\Upsilon(5s)$ resonance. This state can decay into
$B_sB_s$, $B_s^*B_s$ and $B_s^*B_s^*$ where the $B_s^*$ decays to
$B_s\gamma$. As discussed in~\cite{Atwood:2001js} if there are 0 or 2
photons in the final state (i.e. the decay was to $B_sB_s$ or $B_s^*B_s^*$)
then the $B_sB_s$ is in an $L=$ odd state while if there is one photon in
the final state (i.e. $B_s^*B_s$) then the final $B_sB_s$ state is $L=$
even.

The branching ratio to $S_\sigma$ states in the case of $B_s$ is in
principle much larger than in the case of $B_0$.  For instance, the
branching ratio for $B_s\to D_s^+D_s^-$ should be similar to the measured
branching ratio for $B^0\to D^-D^+_s$ which is about 0.8\%. likewise one
can also estimate the branching ratio of $B_s\to J/\psi \eta^{(\prime)}$
at about 0.15\% in addition analogous states such as $D^*_sD^*_s$ etc
should have branching ratios on the order of 0.1\% at least. The
acceptance for such states may be lower than for $\psi K_s$ so we will
assume that ${\cal A}_\sigma{\cal A}_l\approx 0.1$ with a branching ratio
to CP states of about 0.8\%. Using these assumptions, if one had a high
luminosity $\Upsilon(5s)$ machine that was able to produce $10^8$~$B_s$
pairs then $\Delta y\cos\phi=0.7\%$ which would be the same as $\Delta y$
if the Standard Model expectation that $\phi=0$ was correct.  
In~\cite{Atwood:2001js} it was shown that using a method of generalizing
the identification of CP eigenstate decays of $B_s$ to include all states
with a final quark content of $c\bar c s\bar s$ at the expense of the
efficiency of such tagging may allow the determination of $\Delta y$ to a
precision of about $0.28\%$ under the same assumption although in this
case the precision also depends on the value of $y$.

\section{Conclusions}

In summary, we discussed the possibility of time-independent measurements
of lifetime differences in $D$ and $B$ systems. It is important to
reiterate that time-dependent measurements are quite difficult at the
symmetric $e^+e^-$ threshold machines due to the fact that the
pair-produced heavy mesons are almost at rest \cite{Foland:1999wq}. The
techniques described above will provide a {\it time integrated} quantity
that is separately sensitive to the lifetime difference $y$.

This will be particularly useful in the case of $D^0$ where a
charm factory running at the $\psi(3770)$ resonance can yield 
the measurement with precision of $\Delta y\cos\phi\approx 0.5\%$ which
is in the range of some standard model predictions. At a $\Upsilon(5S)$ B
factory with a luminosity sufficient to produce $10^8$ $B_s$ pairs, a
precision of $\Delta y\cos\phi\approx 0.7\%$ should be achievable which is
much smaller that the standard model prediction of 5-15\%.  Thus, even if
only $\sim 10^6$ $B_s$ pairs are produced, precision on the order of the
Standard Model prediction can be obtained.  In the case of the
$\Upsilon(4S)$ B factory with $10^8$ $BB$ pairs, $\Delta y\cos
2\beta\approx 3\%$ which does not probe to the level of the Standard Model
estimate in this case. Yet, a new high luminosity B factory (such as
discussed SuperBaBar) will be able to measure $y$ in $B_d$ system.

\section*{Acknowledgments}
The work of DA was supported by DOE contracts DE-FG02-94ER40817 (ISU) and
DE-AC03-76SF00515 (SLAC) and the hospitality of the SLAC Theory Group. The
work of AP was supported by a Wayne State University Research Grant
Program award. The Authors would also like to thank the hospitality of the
Michigan Center for Theoretical Physics.

\section*{Appendix: Correlated Decays with CP Violation}

In this appendix we provide the expressions for the time integrated decay
of a correlated $H^0-\overline{H^0}$ state to various pairs of final
states using the formalism discussed in the text.

The final states we consider are:

\begin{enumerate}

\item
$S_\pm$, a CP eigenstate such as in $D^0\to K_s\pi^0$ or 
$B_{d(s)} \to J/\psi K_S (\phi)$.
\item
$L^\pm$, a flavor specific semi-leptonic decay to a final state containing
$\ell^\pm$.

\item $G$, a hadronic final state such that both $H^0$ and
$\overline{H^0}$ can decay to it. For example, in charmed mesons, $D^0\to
G^+$ is Cabibbo Favored (CF)  and $D^0\to G^-$ is doubly Cabibbo
suppressed, as in $D^0 \to K^-\pi^+$. This implies that the ratio
of DCS to CA decay rates $R$ is small and the results can be expanded in
terms of this ratio. Alternatively, the ratio of amplitudes can be of
order one in B decay, as in the example of $B_s \to D_s^+ K^-$, so all
powers of R must be kept.

\item $X$ is an inclusive set of all final states. \end{enumerate} It is
now easy to construct all possible combinations of the above final states.
For the case of antisymmetric initial state ($L=$odd), we have for
$\Gamma^{f_1,f_2}_{odd}$

\begin{eqnarray}
\Gamma^{S_\sigma L^\pm}_{odd}
&=&
\left [
2+(1+R_m^{\pm2})x^2-(1-R_m^{\pm2})y^2
\right]
\left [
{
\Gamma_0(S_\sigma)\Gamma_0(L^+)
\over
2 (1-y^2)(1+x^2)
}
\right]
\end{eqnarray}

\begin{eqnarray}
\Gamma^{S_+ S_-}_{odd}
&=&
\bigg[
8R_m^2+(1+R_m^4)(x^2+y^2)
\nonumber\\
&&
+2R_m^2
( (1+2\cos^2\phi)x^2+(1+2\sin^2\phi)y^2)
\bigg]
\times
\nonumber\\
&&
\times
\left[
{
\Gamma_0(S_+)\Gamma_0(S_-)
\over 
2 (1-y^2)(1+x^2)
}
\right]
\end{eqnarray}

\begin{eqnarray}
\Gamma^{S_+ S_+}_{odd}
&=&
\left[
(x^2+y^2)
(\cosh^2\am-\cos^2\phi)
\right]
\left[
\Gamma_0^2(S_\sigma)
\over
(1-y^2)(1+x^2)
\right]
\end{eqnarray}

\begin{eqnarray}
\Gamma^{L^\pm L^\pm}_{odd}
&=&
\left [
R_m^{\mp 2}(x^2+y^2)
\right]
\left [
{
\Gamma_0^2(L^+)
\over
2(1-y^2)(1+x^2)
}
\right]
\end{eqnarray}

\begin{eqnarray}
\Gamma^{L^\pm L^\mp}_{odd}
&=&
\left [
2+x^2-y^2
\right]
\left [
{
\Gamma_0^2(L^+)
\over
2(1-y^2)(1+x^2)
}
\right]
\end{eqnarray}

\begin{eqnarray}
\Gamma^{S_\sigma X}_{odd}
&=&
\bigg [ 
1+y^2\sinh^2\am+x^2\cosh^2\am
\nonumber\\
&&
-
\sigma 
\Big(
y(1+x^2)\cosh\am\cos\phi
+
x(1-y^2)\sinh\am\sin\phi
\Big)
\bigg ]
\times
\nonumber\\
&&
\times
\left[
{
2\Gamma_D\Gamma_0(S_\sigma)
\over
(1-y^2)(1+x^2)
}
\right ]
\end{eqnarray}

\begin{eqnarray}
\Gamma^{L^\sigma X}_{odd}
&=&
\left[ 
1+x^2\cosh^2\am+y^2\sinh^2\am
+\frac{\sigma}{2}(x^2+y^2)\sinh 2\am
\right]
\times
\nonumber\\
&&
\times
\left[
{
2\Gamma_D\Gamma_0(L^+)
\over
(1-y^2)(1+x^2)
}
\right ]
\end{eqnarray}

\begin{eqnarray}
\Gamma^{G^+X}_{odd}
&=&
\bigg [ 
x^2 + y^2 + \left( 1 + R^2 \right) \left( 2 + x^2 - y^2 \right) 
R_m^2 + R^2\left( x^2 + y^2 \right) R_m^4 
\nonumber \\
&&+
2 R x \left(1 - y^2 \right) \sin (\delta  + \phi) R_m
\left(1 - R_m^2 \right)  
\nonumber \\
&&- 
2 R y \left( 1 + x^2 \right) \cos (\delta  + \phi ) R_m
\left(1 + R_m^2 \right) 
\bigg ]
\times
\nonumber\\
&&\times
\left [
{
\Gamma(G^+)\Gamma_D
\over
4(1-y^2)(1+x^2)R_m^2
}
\right]
\end{eqnarray}

\begin{eqnarray}
\Gamma^{GS_\sigma}_{odd}
&=&
\bigg [ 
R_m^2(1+R^2)
+(R^2+R_m^2)(1+R_m^2)x^2
+(R^2-R_m^2)(1-R_m^2)y^2
\nonumber\\
&&
+4rR_m
\Big(
 y^2(\cos\phi\sin\phi\sin\delta -\sin^2\phi\cos\delta)
\nonumber\\
&&
+x^2(\cos\phi\sin\phi\sin\delta +\cos^2\phi\cos\delta)
+\cos\delta
\Big)
\bigg ]
\times
\nonumber\\
&&
\times
\left [
{
2\Gamma(G^+)\Gamma(S_\sigma)
\over
(1-y^2)(1+x^2)R_m^2
}
\right]
\end{eqnarray}

\begin{eqnarray}
\Gamma^{G^\pm L^\pm}_{odd}
&=&
\left [
2R^2+(R_m^{\mp 2}+R^2)x^2+(R_m^{\mp 2}-R^2)y^2 
\right ]
\left [
\Gamma(G)\Gamma(L^+)
\over
2(1-y^2)(1+x^2)
\right]
\end{eqnarray}

\begin{eqnarray}
\Gamma^{G^\pm L^\mp}_{odd}
&=&
\left [
2+(R^2 R_m^{\pm 2}+1)x^2+(R^2 R_m^{\pm 2}-1)y^2 
\right ]
\left [
\Gamma(G)\Gamma(L^+)
\over
2(1-y^2)(1+x^2)
\right]
\end{eqnarray}

\begin{eqnarray} \label{GantiG}
\Gamma^{G^\pm G^\mp}_{odd}
&=&
\biggl[ \left( 1 + R^4 \right) \,R_m^2 \left(-2 - x^2 + y^2 \right)  - 
R^2 \left( x^2 + y^2 \right)  - 
R^2 R_m^4\left( x^2 + y^2 \right)  
\nonumber \\
&+& 
2R^2R_m^2\left[ \left( 2 + x^2 - y^2 \right) \,
\cos 2\delta + \left( x^2 + y^2 \right) \cos 2\phi \right] 
\biggr] 
\times
\nonumber\\
&&
\times
\left [
\Gamma^2(G)
\over
2(y^2-1)(1+x^2)R_m^2
\right]
\end{eqnarray}

\begin{eqnarray} \label{GG}
\Gamma^{G^\pm G^\pm}_{odd}
&=&
\bigg[
(x^2+y^2)
(R^4 R_m^{\pm 2}-2R^2 \cos 2(\delta\pm\phi) + R_m^{\mp 2})
\bigg]
\times
\nonumber\\
&&
\times
\left [
\Gamma^2(G^+)
\over
2(1-y^2)(1+x^2)
\right]
\end{eqnarray}

In the case of charmed mesons, $R\ll 1$. Neglecting possible CP-violating effects 
and taking the ratio of Eqs.~(\ref{GG}) and (\ref{GantiG}) simultaneously
expanding numerator and denominator in $R,x$, and $y$ we reproduce the well-known 
result that DCS/CF interference cancels out in the ratio for 
$L=$odd~\cite{Bigi:1986dp} and gives the result, identical to the semileptonic 
final state, $(x^2+y^2)/2$.

The results for $L=$even are more cumbersome, so we present only a few of 
$\Gamma^{f_1,f_2}_{even}$:

\begin{eqnarray}
\Gamma^{S_\sigma L^\pm}_{even}
&=&
\biggl [
\left( x^2 + y^2 \right)\left(3+\left(x^2 -y^2 \right) + x^2 y^2 \right)
\nonumber \\
&&+
  R_m^{\pm 2}\left( 2 + (1 + x^4) x^2 - \left(1 - 4 x^2 - x^4 \right) y^2 
+
     \left(1 - x^2 \right) y^4 \right)
\nonumber \\
&&- 4 R_m^{\pm 1} \sigma \left(
  {\left( 1 + x^2 \right) }^2 y \cos\phi \mp
  \left(1 - y^2 \right)^2 x \sin\phi \right)
\biggr]
\left [
{
\Gamma_0(S_\sigma)\Gamma_0(L^+)
\over
2 (1-y^2)^2(1+x^2)^2
}
\right]
\end{eqnarray}

%
%
%
%
%

\begin{eqnarray}
\Gamma^{S_+ S_-}_{even}
&=&
\left[
(x^2+y^2)(x^2 +(x^2 - 1)y^2 + 3)
(\cosh 2\am-\cos 2\phi)
\right]
\left[
\Gamma_0^2(S_\sigma)
\over
(1-y^2)^2(1+x^2)^2
\right]
\end{eqnarray}

\begin{eqnarray}
\Gamma^{L^\pm L^\pm}_{even}
&=&
\left [
R_m^{\mp 2}(x^2+y^2)(x^2+(x^2-1)y^2 + 3)
\right]
\left [
{
\Gamma_0^2(L^+)
\over
2(1-y^2)^2(1+x^2)^2
}
\right]
\end{eqnarray}

\begin{eqnarray}
\Gamma^{L^\pm L^\mp}_{even}
&=&
\left [x^4+x^2-(x^2-1)y^4+(x^4+4x^2-1)y^2 + 2)
\right]
\left [
{
\Gamma_0^2(L^+)
\over
2(1-y^2)^2(1+x^2)^2
}
\right]
\end{eqnarray}

Taking the ratios of the decay rates presented above, one can easily generalize 
the results of \cite{Bigi:1986dp} to the case of CP-nonconservation.



\end{document}